\documentclass[twocolumn,showpacs,preprintnumbers]{revtex4}
\usepackage{bm}
\usepackage{amssymb}
\usepackage{amsmath}
\usepackage{graphicx}
\begin{document}
\title{Neutrino Scattering in Heterogeneous Supernova Plasmas}
\author{O.L Caballero}\email{lcaballe@indiana.edu}
\affiliation{Department of Physics and Nuclear Theory Center,
             Indiana University, Bloomington, IN 47405}
\author{C.J. Horowitz}\email{horowit@indiana.edu}
\affiliation{Department of Physics and Nuclear Theory Center,
             Indiana University, Bloomington, IN 47405}
\author{D. K. Berry\footnote { e-mail: dkberry@indiana.edu}}
\affiliation{University Information Technology Services,
             Indiana University, Bloomington, IN 47408}

\date{\today}
\begin{abstract}
Neutrinos in core collapse supernovae are likely trapped by neutrino-nucleus elastic
scattering.  Using molecular dynamics simulations, we calculate neutrino mean free
paths and ion-ion correlation functions for heterogeneous plasmas.   Mean free paths
are systematically shorter in plasmas containing a mixture of ions compared to a plasma
composed of a single ion species.  This is because neutrinos can scatter from concentration
fluctuations.  The dynamical response function of a heterogeneous plasma is found to
have an extra peak at low energies describing the diffusion of concentration fluctuations.
Our exact molecular dynamics results for the static structure factor reduce to the
Debye Huckel approximation, but only in the limit of very low momentum transfers.  
\end{abstract}
\smallskip
\maketitle

\section{Introduction}
Core collapse supernovae are extraordinarily energetic explosions where 99\% of the
energy released is radiated in neutrinos \cite{ref1,ref1a}.  This is because weakly
interacting neutrinos are the only known particles that can diffuse quickly out of
the dense stellar core.  When the density of the core reaches $\approx 10^{12}$ g/cm$^3$,
neutrino nucleus elastic scattering is thought to temporarily trap the neutrinos \cite{ref2}.
The neutrinos then provide a degeneracy pressure that helps support the star.  If the
neutrinos were not trapped, the star could collapse to a black hole without a large
supernova explosion.  Thus, the dynamics of a supernova is sensitive to how neutrinos
interact in dense matter.  For example, changes in the electron capture rate on nuclei
could change the sensitive balance between electron Fermi pressure and gravity
\cite{ecap1,ecap2}.  However electron capture can only proceed if the produced neutrinos
are not Pauli blocked. 

The neutrino wave length is comparable to the spacing between ions.  Therefore, ion-ion
correlations can significantly modify neutrino-nucleus scattering cross sections.  An
ion in a plasma will be surrounded by a screening cloud of other ions and perhaps
electrons.  Neutrino interactions with this cloud will, in general, reduce neutrino-nucleus
cross sections and increase the neutrino mean free path, see for example \cite{Ho97}.
Most calculations of this screening assume a one component plasma composed of a single
species of ion, see for example \cite{itoh}.  A first attempt was made in ref. \cite{janka}
to consider scattering from a plasma including both alpha particles and heavy nuclei.
However, this was based on a simple prescription to describe the mixture.

If the plasma has a mixture of species with different ratios of weak to electromagnetic
charges then the screening of neutrino reactions can be very different.  Neutrinos scatter
from fluctuations in the weak charge density.  Fluctuations in composition, that change
the weak charge density but do not change the (electric) charge density, will feel no
electric restoring force.  Therefore, these fluctuations may not be screened and can lead
to a large change in the neutrino cross section.  This was first discussed in connection
with ion and electron screening in refs. \cite{escreen1,escreen2}.  Recently  Sawyer
\cite{sawyer} argued that composition fluctuations for a mixture of ions with different
ratios of neutron number $N$ to charge $Z$ can increase the cross section.  However
Sawyer used the Debye Huckel approximation which is only valid at very low momentum transfers.   

In this paper we use molecular dynamics simulations to accurately calculate neutrino
scattering cross sections and mean free paths from a dense plasma composed of mixtures
of different ions.  Section II discusses the mixture of ions expected in supernova
plasmas and presents our neutrino scattering formalism, section III describes our
simulations, section IV our results and we conclude in section V.   

\section{Formalism}
\label{sec:formalism}

We are interested in neutrino scattering from a dense plasma during the infall phase
of a supernova.  As the density and temperature rise the rate of nuclear reactions
increases dramatically.  This should bring the composition of the plasma into nuclear
statistical equilibrium (NSE).  In NSE all nuclei are in chemical equilibrium so the
abundance of a given state is determined by its binding energy and entropy.  We first
discuss this NSE composition and then describe how to calculate neutrino scattering
from this heterogeneous mixture.

\subsection{Composition}
\label{subsec:composition}  

The composition of a plasma in nuclear statistical equilibrium depends on the density,
temperature and proton fraction.  For given conditions, we expect the most abundant
heavy isotope to have a binding energy and entropy that leads to a large free energy.
However nuclear reactions can quickly add or subtract nucleons to this most abundant
species.  Therefore, we expect a distribution of isotopes near the most abundant one.
In addition, entropy can favor some light isotopes such as alpha particles and free
neutrons.  There have been many statistical models that calculate the NSE composition,
see for example \cite{botvina}.  Typically, these statistical approaches incorporate
detailed binding energy models but only include the interactions between nuclei
approximately.  
            
Alternatively, a semi-classical microscopic model described in ref. \cite{pasta, pasta2}
includes the strong interactions between nucleons in different nuclei in an identical
fashion to the interactions between nucleons in a given nucleus.  This allows the model
to describe not only a plasma of isolated ions but higher density matter where the
nuclei merge together to form complex pasta shapes and then uniform nuclear matter. 

However, this simple model may not reproduce the binding energy of individual nuclei
as well as statistical models.  Although the force has been fit to reproduce the
binding energy of nuclear matter, the model does not include pairing and shell structure.
Comparing the microscopic and statistical models provides some estimate of the possible
range in compositions that could be expected.

In the microscopic model, the location of each nucleon is followed in a molecular
dynamics simulation.  For example, in ref. \cite{pasta2} trajectories for 40,000
nucleons have been calculated by integrating Newton's laws.  At any instant in time,
the configuration of nucleons is divided into nuclei using the following simple algorithm.
A nucleon is said to belong to a given nucleus if it is within a cutoff radius $R_{cut}=3$
fm of at least one other nucleon in the nucleus.  This algorithm uniquely divides a
given configuration of many nucleons into a collection of nuclei.  The resulting
distribution of nuclei will be presented in section \ref{sec:results}.  However,
first we present our neutrino scattering formalism.

\subsection{Neutrino Scattering}
\label{subsec:nuscat}
In this section we describe how neutrino scattering is modified by ion-ion correlations.
The free neutrino-nucleus elastic scattering cross section is 
\begin{equation}
d\sigma_0/d\Omega = \frac{G^2C^2E^2_\nu(1+\cos\theta)}{4\pi^2}.
\label{sigma0}
\end{equation}
Here $G$ is the Fermi constant, $E_\nu$ is the neutrino energy, $\theta$ the scattering
angle and $C$ is the total weak charge of a nucleus with charge $Z$ and neutron number $N$,
\begin{equation}
C=-2Z\sin^2\Theta_W+(Z-N)/2,
\end{equation} 
with a Weinberg angle of $\sin^2\Theta_W=0.223$.  Note that $\sin^2\Theta_W\approx 0.25$.
In the following we approximate $C\approx -N/2$.  For a mixture of ions we will use,
\begin{equation}
C=-\frac{1}{2} \langle N \rangle,
\label{weakcharge}
\end{equation}
where the average neutron number $\langle N \rangle$ is, 
\begin{equation}
\langle N \rangle = \frac{1}{N_{ion}} \sum_{i=1}^{N_{ion}} N_i,
\end{equation}
for a system of $N_{ion}$ total ions where the ith ion has neutron number $N_i$.

Ion correlations can be taken into account by multiplying $d\sigma_0/d\Omega$ by the
static structure factor $S(q)$~\cite{No64},
\begin{equation}
d\sigma/d\Omega=d\sigma_0/d\Omega\, S(q).
\label{sigma}
\end{equation}
Here $q$ is the momentum transfer and $\sigma_0$ is the free cross section.
The static structure factor adds coherently the contributions for neutrino scattering
from different nuclei, including the relative phases, and can be calculated from,
\begin{equation}
S(\mathbf{q})=\frac{1}{N_{ion}}\left(\langle\hat{\rho}^\dagger(\mathbf{q})\hat{\rho}
(\mathbf{q})\rangle-\left|\langle\hat{\rho}(\mathbf{q})\rangle\right|^2\right),
\end{equation}
with $\hat{\rho}(\mathbf{q})$ the density operator is given by,
\begin{equation}
  \hat{\rho}(\mathbf{q})=\displaystyle\sum_{i=1}^{N_{ion}} \frac{N_i}{\langle N \rangle}
                         \exp(i\mathbf{q}\cdot{\mathbf{r}}_i).
  \label{rhohat}
\end{equation}
Note the choice of the normalization $\langle N \rangle$ in Eqs. (\ref{weakcharge},\ref{rhohat})
is a somewhat arbitrary convention.  We make this choice because one often approximates
a mixed system with a single species where all the $N_i=\langle N \rangle$.  Alternatively
one could replace $\langle N \rangle$ by $\langle N^2 \rangle^{1/2}$ in both Eqs.
(\ref{weakcharge}) and (\ref{rhohat}) with no change in the cross section in Eq. (\ref{sigma}).
With this second choice $S(q)\rightarrow 1$ at high $q$.

The transport mean-free path $\lambda$ is inversely proportional to the transport
cross section $\sigma^t$, $\lambda=1/\rho_i\sigma^t$ with $\rho_i$ the ion density.
In a medium, the transport cross section $\sigma^t$ can be obtained by multiplying
the free transport cross section
\begin{equation}
  \sigma_0^t=\int d \Omega (1-\cos\theta) d\sigma_0/d\Omega= \frac{2}{3}G^2C^2E_\nu^2/\pi
\end{equation} 
by $\langle S\rangle$,
\begin{equation}
  \sigma^t=\sigma_0^t\langle S\rangle,
  \label{crosst}
\end{equation}
where $\langle S\rangle$ is the angle average of $S(q)$~\cite{Ho97},
\begin{equation}
\langle S(E_\nu,\rho,T)\rangle=\frac{3}{4}\int_{-1}^{1}d\cos\theta(1+\cos\theta)
(1-\cos\theta)S(q(\theta)).
\end{equation}
Here $q(\theta)^2=2E_\nu^2(1-\cos\theta)$ and the factor of $(1+\cos\theta)$ comes from
angular dependence of the free cross section, Eq. (\ref{sigma0}).

\subsection{Debye Huckel Approximation}
\label{subsec:debyehuckel}

At very low momentum transfers $q$ the simple Debye Huckel approximation is valid \cite{Fe71}.
This provides insight into our full molecular dynamics simulation results.  The static
structure factor in the Debye Huckel approximation $S_q^{DH}$ is given by \cite{sawyer},
\begin{equation}
  S_q^{DH}=\frac{T}{\rho_i \langle N \rangle^2} \sum_{i,j=1}^{N_{ion}} N_i N_j K_{ij}.
\end{equation}
Here $T$ is the temperature and
\begin{equation}
  K_{ij}=\Pi_{ij} -\frac{\bigl(\sum_k Z_k \Pi_{ik} \bigr)\bigl(e^2\sum_l Z_l \Pi_{lj}\bigr)}
                        {q^2 + q_e^2 + e^2 \sum_{m,n} Z_m Z_n \Pi_{mn}},
\end{equation}
where electron screening is describe by $q_e^2 = (4\alpha/\pi) k_F^2$.  In the Debye
Huckel approximation,
\begin{equation}
  \Pi_{ij}=\frac{\delta_{ij} n_i}{T} = \frac{\delta_{ij}}{VT},
\end{equation}
where we assume the density of the ith ion is $n_i=1/V$ with $V$ the system volume.
The total ion density is $\rho_i=N_{ion}/V$.  Using this expression for $\Pi_{ij}$
reduces $K_{ij}$ to,
\begin{equation}
  K_{ij}=\frac{1}{VT}\Bigl\{\delta_{ij}-\frac{e^2}{VT}\, \frac{Z_i Z_j}{q^2 + \kappa^2}\Bigr\},
\end{equation}
with $\kappa^2=\kappa_{ion}^2+q_e^2$ and
\begin{equation}
  \kappa_{ion}^2=\frac{e^2}{VT}\sum_i Z_i^2.
\end{equation}
The final result is 
\begin{equation}
  S_q^{DH} = \frac{\langle Z^2 \rangle}{\langle N \rangle^2}\Bigl\{ \langle \lambda^2 \rangle
             - \frac{\langle \lambda \rangle^2 \kappa_{ion}^2}{q^2+\kappa^2}\Bigr\},
\label{SDH}
\end{equation}
where,
\begin{equation}
   \langle \lambda^k \rangle = \frac{\sum_i (\frac{N_i}{Z_i})^k Z_i^2}{\sum_i Z_i^2},
\end{equation}
for $k=1,2$ and,
\begin{equation}
   \langle Z^2 \rangle = \frac{1}{N_{ion}} \sum_i Z_i^2.
\end{equation}
In the $q=0$ limit, and neglecting electron screening $q_e^2=0$, we have
\begin{equation}
   S_0^{DH}=\frac{\langle Z^2 \rangle}{\langle N \rangle^2}\, \, \bigl(\Delta \lambda\bigr)^2.
\end{equation}
Thus the static structure factor depends on the dispersion in the ratio of weak to
electromagnetic charge, with
     $(\Delta \lambda)^2=\langle \lambda^2 \rangle - \langle \lambda \rangle^2$.
Alternatively for a single component plasma, we have
\begin{equation}
   S_q^{DH}=\frac{q^2+q_e^2}{q^2 + \kappa^2}.
\end{equation}
We compare our full simulation results to the Debye Huckel expression, Eq. (\ref{SDH}),
in Section \ref{sec:results}.

\section{Simulations}
\label{sec:simulations}

In order to calculate the static struture factor, for arbitrary momentum transfer
$q$, we perform molecular dynamics simulations using the Verlet algorithm~\cite{Er97}.
For the conditions we consider, the thermal deBrogile wavelength of the ions is much
less than the inter-ion spacing so we assume that the ions behave classically. We are
interested in momentum transfers much less than the electron Fermi momentum $q<<k_F$.
Therefore, we can describe the interaction between ions with a Yukawa potential~\cite{Fe71}:
\begin{equation}
   V(i,j)=\frac{Z_i Z_je^2}{4\pi r_{ij}}e^{-r_{ij}/\lambda_e},
\end{equation}
where $r_{ij}$ is the distance between a pair of ions.  The electron screening length is
$\lambda_e=\pi/(ek_F)$ where the electron Fermi momentum is
$k_F=(\rho_e 3\pi^2)^{1/3}$, the electron density is $\rho_e=\langle Z \rangle \rho_i$, and $e^2=4\pi\alpha$.

Periodic boundary conditions are used to minimize finite size effects. The distance
between ions $r_{ij}$ is then given in terms of the coordinates $x, y$ and $z$ of the $i$th
and $j$th particles in the form:
\begin{equation}
   r_{ij}=\sqrt{[x_i-x_j]^2+[y_i-y_j]^2+[z_i-z_j]^2}.
\end{equation}
Then, the distance used for the periodic boundary condition is:
\begin{equation}
   [l]\equiv\min{(|l|,L-|l|)}.
\end{equation}
Here $L$ is the side of the cubic box $V$ containing the $N_{ion}$ ions, $L=V^{1/3}$.
For ion density $\rho_i$ the box volume is $V=N_{ion}/\rho_i$.


We used three different Fortran molecular dynamics programs to do our simulations.  The programs are indicated by the author's initials.  All the codes use the direct particle-particle method to calcluate interactions, where one directly sums the $\frac{1}{2}N(N-1)$ interactions among $N$ ions. Thus the amount of work to do a simulation increases as the square of the number of particles.

The CH and LC codes are serial codes suitable for doing small problems. They were
used for MD simulations of 1000 to 4000 ions. The larger 10000 and 40000 ion runs
however were too much work for a serial code, so these were done with a parallel
program, DB. The 40000 ion mixture run was particularly time-consuming, both because
of the 100 fold increase in work per time step over the 4000 ion run, and because it
had to be run longer in order to capture the diffusion of the weak charge fluctuation
across the larger simulation box. DB uses the MPI (Message Passing Interface) library
to pass messages among processes. The $N$ ions are partitioned into $p$ sets, where
$p$ is the number of MPI processes, and each process is tasked with calculating the
forces on its set of $N/p$ particles due to all $N$. This makes for $\frac{1}{2}N(N-1)/p$
interactions per process per time step. The main communication needed is for all
processes to share their coordinates with each other, what is called an ``allgather''.
This can result in a high communication overhead, which is relatively less costly for
large problem sizes. After swapping coordinates, the force calculation and integration
of Newton's equations can proceed without communication. The force calculation in each
MPI process is additionally parallelized with OpenMP, which is a set of compiler
directives used primarily for parallelizing DO loops on multiprocessor shared memory
machines.

Run DB1 was done on a 36 node distributed memory parallel computer at Indiana University,
the AVIDD-O machine, where each node board has two AMD Opteron processors (Advanced
Micro Devices).  Each node ran one MPI process consisting of two OpenMP threads. We
started with the 40000 ions uniformly and randomly distributed in a simulation box of
edge length $L=822.8$ fm, and with velocities distributed according to a Boltzmann
distribution at temperature $T=1.00$ MeV. We brought the system to equilibrium during
the first 7000 warmup steps, with a time step $\Delta t=25$ fm/c, and then ran a further
252500 measurement steps, with $\Delta t=50$ fm/c. Every tenth configuration was written
to disk. These were used for calculating $S(q,w)$ and $S(q)$.  The velocity-Verlet
algorithm was used to integrate Newton's equations. As this algorithm conserves total
energy, we rescaled velocities every 1250 time steps in order to maintain the temperature
at $T=1.00 MeV$. We used 4 Opteron nodes (8 processors) for the warmup steps, 8 nodes
(16 processors) for the first 65000 measurement time steps, and 10 nodes (20 processors)
for the remaining 187500. The program did not have perfect linear speedup with the number
of nodes. A 40000 particle problem is actually too small to run efficiently on this
number of processors, due to communication overhead. Speedup becomes more linear as the
problem size is increased.  The full DB1 run consisted of 259500 MD time steps, and took
793 hours spread out over 8 weeks.

We did the 10000 pure ion run DB2 on one dual processor PowerPC970 node in Indiana
University's 42 node IBM JS20 system. For this run we compiled the code to use only
the OpenMP capability, as the 2 processors share the board memory. We again started
with the ions uniformly and randomly distributed in a simulation box of edge length
$L=518.4$ fm, and with velocities distributed according to a Boltzmann distribution
at $T=1.00$ MeV. We did 8000 warmup steps at $\Delta t=25$ fm/c, and 24000 more at
$\Delta t=50$ fm/c. We then did 100000 measurement steps at $\Delta t=50$ fm/c, again
writing out every tenth configuration for analysis. Velocities were rescaled every
1250 time steps to maintain the temperature at $T=1.00$ MeV. The full DB2 run consisted
of 132000 MD time steps and took 346 hours spread out over 16 days. We did run DB2
almost simultaneously with DB1, thus running on two processors caused no extra delay.

Finally, we measure the static structure factor $S(q)$ for valuse of $\bf q$,
\begin{equation}
\{ q_x, q_y, q_z \} = \frac{2\pi}{L} (n_x, n_y, n_z)\, .
\end{equation} 
To minimize finite size effects we chose $n_x$, $n_y$, $n_z$, as integers.  We average over directions of $\bf q$ to improve the statistics.  Statistical error bars for codes CH and DB are estimated from the dependence of $S(q)$ on the direction of $\bf q$.  Statistical error bars for code LC are estimated by dividing the measurement time into a number of separate groups and looking at the variance of $S(q)$ for different groups.

\section{Results}
\label{sec:results}
In this section we present results for the static structure factor from full molecular
dynamics simulations and compare to the simple Debye Huckel approximation.  However
first we present results for the plasma composition based either on a simple microscopic
model or on a statistical model.

\subsection{Composition}
\label{subsec:compresults}
 
In this subsection we present results for the distribution of charge $Z_i$ and neutron
number $N_i$ to be used in our molecular dynamics simulations.  We start with conditions
considered in ref. \cite{pasta2}:  a density of 0.01 nucleons per $fm^3$ or
$1.66\times 10^{13}$ g/cm$^3$, a temperature of $T=1$ MeV and a proton fraction
$Y_p=0.2$.  Reference \cite{pasta2} performed a molecular dynamics simulation in the
nucleon coordinates with a simple short range nuclear plus Coulomb interaction.  The
nuclear interaction was fit to reproduce the saturation density and binding energy of
nuclear matter.

\begin{figure}[ht]
\begin{center}
\includegraphics[width=2.75in,angle=270,clip=true] {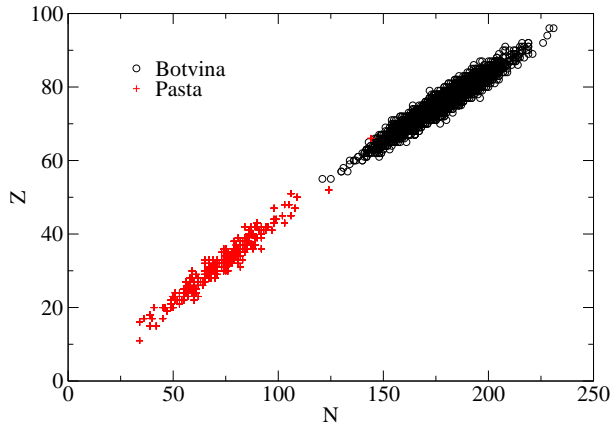}
\caption{Distribution of charge $Z$ versus neutron number $N$ for ions at a temperature
  of $T=1$ MeV and proton fraction $Y_p=0.2$.  The microscopic Pasta results are from
  ref. \cite{pasta2} at a density of $1.66\times 10^{13}$ g/cm$^3$ while statistical
  model results are from Botvina et al. \cite{botvina} at a density of $10^{13}$ g/cm$^3$.} 
\label{Fig1}
\end{center}
\end{figure}

The final coordinates of a simulation with 40000 nucleons are divided into nuclei by
assuming a nucleon belongs to a nucleus if it is within 3 fm of at least one other
nucleon in the nucleus.  This algorithm produces a collection of about 300 medium mass
nuclei with $\langle A \rangle\approx 100$ shown in Fig. \ref{Fig1}.  In addition there
are a few light nuclei with mass $A<10$ (not shown) and about 10000 free neutrons.      

We are interested in the effects of ion-ion correlations so we neglect the free
neutrons.  Neutrino interactions with these free neutrons are not expected to have
large correlation effects.  In addition, we also neglect the light nuclei.  The
light nuclei have small weak and electromagnetic charges, and are expected to play
only a small role in neutrino scattering.  Furthermore, the light nuclei have larger
thermal velocities that require a smaller molecular dynamics time step which slows
down the molecular dynamics simulations.  

\begin{figure}[ht]
\begin{center}
\includegraphics[width=2.75in,angle=270,clip=true] {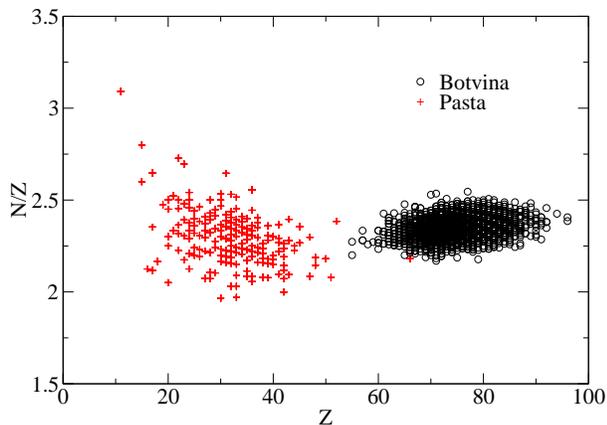}
\caption{Ratio of neutron number to charge $N/Z$ versus $Z$ for ions at a temperature
  of $T=1$ MeV and proton fraction $Y_p=0.2$.  The microscopic Pasta results are from
  ref. \cite{pasta2} at a density of $1.66\times 10^{13}$ g/cm$^3$ while statistical
  model results are from Botvina et al. \cite{botvina} at a density of $10^{13}$ g/cm$^3$.} 
\label{Fig2}
\end{center}
\end{figure}

We compare these microscopic results to the statistical model of Botvina et al.
\cite{botvina} which we also show in Fig. \ref{Fig1}.  For a slightly lower density
$10^{13}$ g/cm$^3$ and the same $Y_p$ and $T$, the statistical model yields significantly
larger nuclei $\langle A \rangle \approx 300$.  Average $\langle A \rangle$ and
$\langle Z \rangle$ are collected in Table \ref{tableone} for these two models.
The dispersion in the ratio of weak to electromagnetic charge $N/Z$ is important
for neutrino scattering.  This ratio is plotted in Fig. \ref{Fig2} and the dispersion
is more similar for the two models although somewhat larger for the microscopic model.       

\begin{table}
\caption{Results for ion density $\rho_i$, average mass number $\langle A \rangle$,
   average charge $\langle Z \rangle$, and mass fraction of free neutrons $X_n$ for
   microscopic \cite{pasta2} and statistical \cite{botvina} models.} 
\begin{tabular}{lllllll}
Model & $\rho$ (gm/cm$^3$) & $Y_p$ & $\rho_i$ (fm$^{-3}$) & $\langle A \rangle$ & $\langle Z \rangle$ & $X_n$ \\
Microscopic & $1.66\times 10^{13}$ & 0.2 & $7.18\times 10^{-5}$ & 105.2 & 32.1 & 0.25 \\
Statistical & $10^{13}$ & 0.2 & $1.60\times 10^{-5}$ & 328.6 & 75.7 & 0.33 \\
Statistical & $10^{12}$ & 0.5 & $1.06\times 10^{-5}$ & 56.8 & 28.3 & $\approx 0$ \\ 
\end{tabular} 
\label{tableone}
\end{table}

\begin{figure}[ht]
\begin{center}
\includegraphics[width=2.75in,angle=270,clip=true] {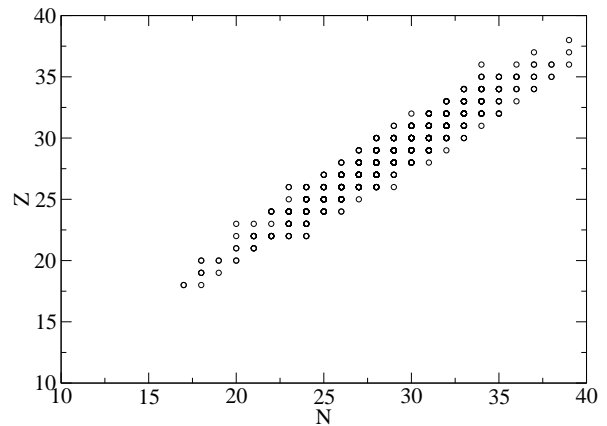}
\caption{Distribution of charge $Z$ versus neutron number $N$ for ions in a statistical
  model \cite{botvina} at a density of $10^{12}$ g/cm$^3$, temperature of $T=1$ MeV and
  proton fraction $Y_p=0.5$.  } 
\label{Fig3}
\end{center}
\end{figure}

\begin{figure}[ht]
\begin{center}
\includegraphics[width=2.75in,angle=270,clip=true] {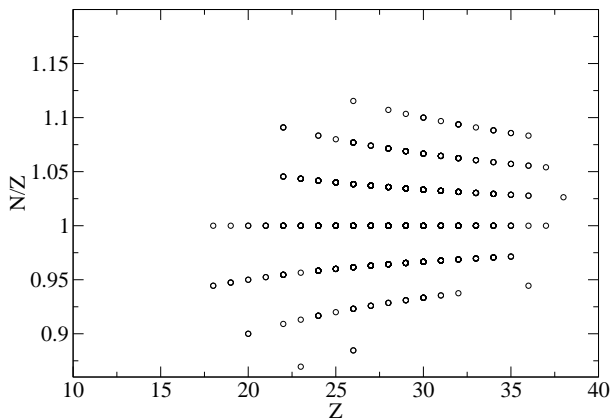}
\caption{Ratio of neutron number to charge $N/Z$ versus $Z$ for ions in a statistical
  model \cite{botvina} at a density of $10^{12}$ g/cm$^3$, temperature of $T=1$ MeV
  and proton fraction $Y_p=0.5$.} 
\label{Fig4}
\end{center}
\end{figure}

In a core collapse supernova the proton fraction starts out just below $Y_p=0.5$
and decreases with time due to electron capture.  Therefore, in Figs. \ref{Fig3},
\ref{Fig4} we show statistical model results for $Y_p=1/2$, $T=1$ MeV and a density
of $10^{12}$ g/cm$^3$.  Note that the microscopic model has not been simulated for
$Y_p=1/2$.  The dispersion in $N/Z$ is smaller for $Y_p=1/2$ than for $Y_p=0.2$.

\subsection{Static Structure Factor}
\label{subsec:S_q}

\begin{figure}[ht]
\begin{center}
\includegraphics[width=2.75in,angle=270,clip=true] {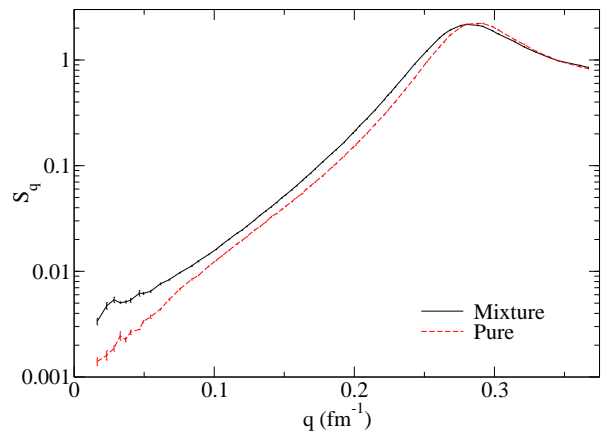}
\caption{ Static structure factor $S(q)$ versus momentum transfer $q$ for a system
  at a density of $1.66\times 10^{13}$ g/cm$^3$, $Y_p=0.2$, and $T=1$ MeV.  The
  solid curve uses a distribution of ions from a microscopic model (CH1) while the dashed
  line assumes a single component plasma (CH2). Results are from molecular dynamics
  simulations using 4000 ions.  Note the log scale. } 
\label{Fig5}
\end{center}
\end{figure}

\begin{figure}[ht]
\begin{center}
\includegraphics[width=2.75in,angle=270,clip=true] {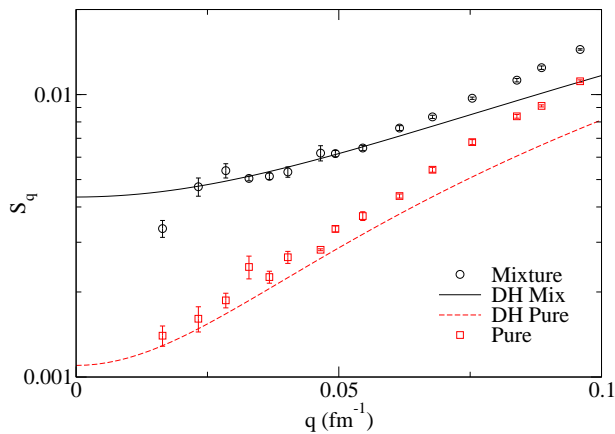}
\caption{ Detail of the static structure factor $S(q)$ versus momentum transfer $q$
  at low $q$ from Fig. \ref{Fig5}.  Debye Huckel results, Eq. (\ref{SDH}) for a
  mixture of ions are shown as the solid line while the dashed line shows results
  for a pure system.   Note the log scale. }
\label{Fig6}
\end{center}
\end{figure}

In this subsection we present molecular dynamics simulation results for the static
structure factor $S(q)$. We ran two simulations with 4000 ions. The first, labeled CH1 in
Table \ref{tabletwo}, was based on the mixture in Fig. \ref{Fig1} from the pasta results of \cite{pasta2}.
The second, labeled CH2, was for a pure system assuming a single component plasma where
each ion has neutron number $N_i=\langle N \rangle$ and charge $Z_i=\langle Z \rangle$ equal
to the average values of the mixture. Both simulations were done at
density $1.66\times 10^{13}$ g/cm$^3$, proton fraction $Y_p=0.2$, and temperature
$T=1$ MeV.
For these conditions the ion density is $\rho_i=7.18\times 10^{-5}$ fm$^{-3}$, see
Table \ref{tableone}.
The mixture simulation was started from random coordinates and warmed up for a time
of 500000 fm/c. The system was then evolved for a further $1.6\times 10^6$ fm/c during
which $S(q)$ was measured from configurations written to disk every 500 fm/c, see Table
\ref{tabletwo}. The pure system was also started from random coordinates, and warmed up for a
time of $1.6\times 10^6$ fm/c. It was then evolved for $1.0\times 10^6$ fm/c, with $S(q)$
again measured from configurations written out every 500 fm/c. $S(q)$ for these runs are
compared in Fig. \ref{Fig5}.
In Fig. \ref{Fig6} we compare them for small $q$, as well as with the Debye
Huckel results of Eq. (\ref{SDH}).

\begin{table}
\caption{Simulation runs: the composition is from a microscopic model (Pasta) \cite{pasta2}, a statistical model of Botvina (Bot.) \cite{botvina} or a single (pure) species, the time step is $\Delta t$, while $T_W$ is the warm up time and $T_M$ the measurement time, and the potential energy per ion is $\langle V \rangle$ } 
\begin{tabular}{llllllll}
Run & $N_{ion}$ & Comp. & $\Delta t$  & $\rho_i (10^{-5}$  & $T_W (10^5$  & $T_M (10^6$  & $\langle V \rangle$ \\
     &           &        &  fm/c & fm$^{-3})$ & fm/c) &  fm/c) & (MeV) \\   
CH1 & 4000 & Pasta & 50  & $7.18$ & $5$ & $1.6$ & 367.467(1) \\ 
CH2 & 4000 & Pure & 50 &  $7.18$ & $1.6$ & $1$ &  368.94(1) \\
CH3 & 1000 & Bot. & 100 & $1.06$ & $4.1$ & $8$ & 162.952(2) \\
CH4 & 1000 & Pure & 100 & $1.06$ & $2.4$ & $8$ & 161.990(3) \\
LC1 & 1000 & Pasta & 56 & $7.18$ & $2.8$ & $1.1$ & 357.65(9)\\
LC2 & 1000 & Pure & 56 & $7.18$ & $2.8$ & $1.1$ & 359.08(8)\\
DB1 & 40000 & Pasta& 50 & $7.18$ & $1.8$ & $13$ & 367.712(1) \\
DB2 & 10000 & Pure& 50 & $7.18$ & $14$ & $5$ & 368.151(1) \\

\end{tabular} 
\label{tabletwo}
\end{table}

The first thing we note from these figures is that $S(q)$ for the pure system
is systematically smaller than $S(q)$ for the mixture.  This is because the mixture can
have additional fluctuations in the weak charge density from interchanging isotopes
that do not have corresponding fluctuations in the charge density, see below.       

Second, Figure \ref{Fig6} shows good agreement between the MD simulation results for $S(q)$
and the Debye Huckel approximation $S_q^{DH}$, Eq. \ref{SDH}, for $q$ between 0.025
and 0.055 fm$^{-1}$.  For $q>0.055$ fm$^{-1}$, $S_q^{DH}$ is smaller than $S(q)$ from
MD.  Therefore, $S_q^{DH}$  is only valid for small $q$.  The lowest $q$ MD result
in Fig. \ref{Fig6} corresponds to $q=q_{min}=2\pi/L$ where $L^3$ is the simulation
volume.  There may be a systematic error for this point associated with the finite
measurement time of $1.6\times 10^6$ fm/c for the MD calculation of $S(q)$.  It may
take a long time for fluctuations in the weak charge to diffuse across the system and
this may require a long measurement time to obtain an accurate $S(q)$ at low $q$.
We discuss this further in subsection \ref{subsec:finitesize}.
   
We also performed MD simulations comparing the statistical mixture of Botvina against
a pure system. These are labeled CH3 and CH4 in Table \ref{tabletwo}, and involve just 1000 ions.
Rather using the Botvina distribution
shown in Fig. \ref{Fig1}, however, where $\rho=10^{13}$ g/cm$^3$ and $Y_p=0.2$,
we used the mixture of Figs. \ref{Fig3}, \ref{Fig4}, where $\rho=10^{12}$ g/cm$^3$ and
$Y_p=0.5$.  We kept the temperature at $T=1$ MeV.
The reason is that the denser system may be a solid.
The ratio of a typical Coulomb to thermal
energy is $\Gamma=Z^2\alpha/aT$ where the ion sphere radius is $a=(3/4\pi\rho_i)^{1/3}$.
Assuming a pure system with $Z=\langle Z \rangle = 75.7$ we have $\Gamma=335$.  A pure
one component plasma is expected to be a solid for $\Gamma > 180$.  Note, the behavior
of a mixture of ions could be somewhat different from a pure system.  However, given
the very large value of $\Gamma$ the mixture could still solidify.    

\begin{figure}[ht]
\begin{center}
\includegraphics[width=2.75in,angle=270,clip=true] {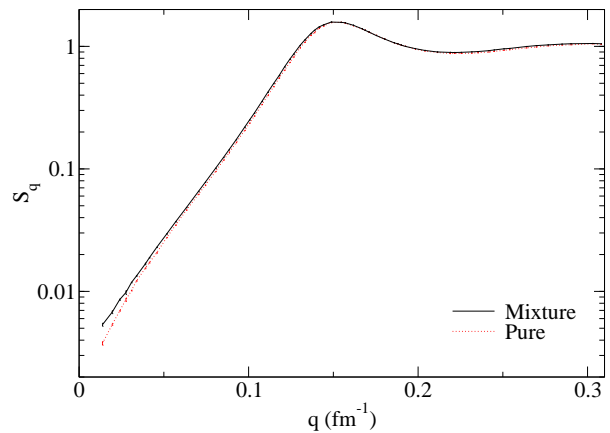}
\caption{Static structure factor $S(q)$ versus momentum transfer $q$ for a system at
   a density of $10^{12}$ g/cm$^3$, $Y_p=0.5$, and $T=1$ MeV.  The solid curve uses
   a distribution of ions from a statistical model (run CH3) while the dashed line assumes a
   single component plasma (run CH4). Results are from molecular dynamics simulations using 1000
   ions.  Note the log scale.} 
\label{Fig7}
\end{center}
\end{figure}

\begin{figure}[ht]
\begin{center}
\includegraphics[width=2.75in,angle=270,clip=true] {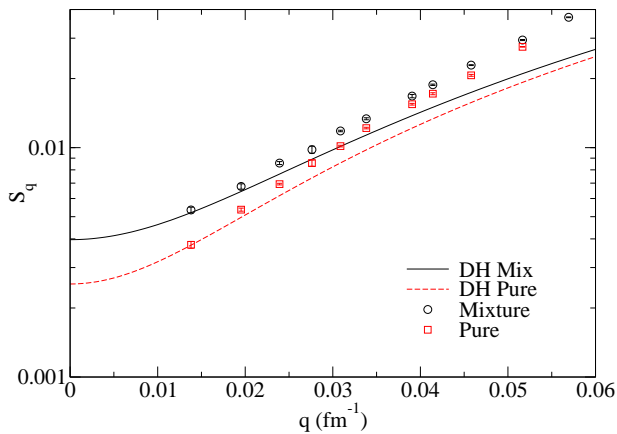}
\caption{Detail of the static structure factor $S(q)$ versus momentum transfer $q$
   at low $q$ from Fig. \ref{Fig7}.  The circles are results for a mixture of ions
   while the squares are for a pure system.  Debye Huckel results, Eq. (\ref{SDH})
   for a mixture of ions are shown as the solid line while the dotted line shows
   results for a pure system.   Note the log scale.} 
\label{Fig8}
\end{center}
\end{figure}

The $S(q)$ curves of runs CH3 and CH4 are shown in Figs. \ref{Fig7} and \ref{Fig8}.
Now there is agreement between MD and Debye Huckel results near
$q\approx 0.015$ fm$^{-1}$.  For $Y_p=0.5$ the spread in $N/Z$ for the ions is smaller
than at $Y_p=0.2$.  As a result, $S(q)$ for the
mixture is enhanced over $S(q)$ for the pure system by a smaller amount than at $Y_p=0.2$.
These results for $S(q)$ will be used in subsection \ref{subsec:mfp} to calculate
neutrino mean free paths.  However, first we present results for the dynamical response
function to gain insight into the difference between $S(q)$ for mixtures and $S(q)$
for a pure system.

\subsection{Dynamical Response Function}
\label{subsec:dynamical}
\begin{figure}[ht]
\begin{center}
\includegraphics[width=2.75in,angle=270,clip=true] {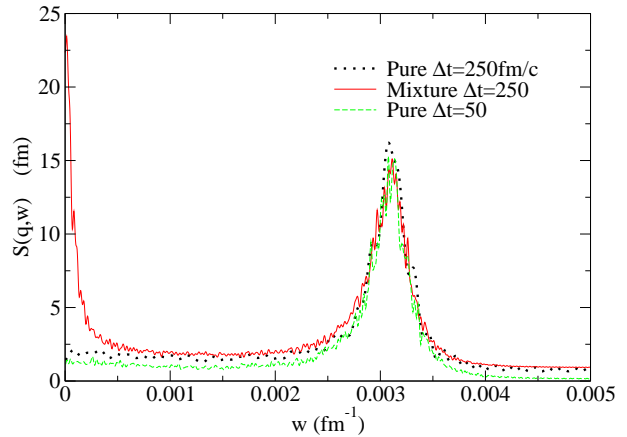}
\caption{Dynamical response function $S(q,w)$ versus energy transfer $w$ for
   $q=0.116$ fm$^{-1}$ at a density of $1.66\times 10^{13}$ g/cm$^3$, $Y_p=0.2$
   and $T=1$ MeV, from MD simulations with 4000 ions.  The solid curve assumes a
   mix of ions from the microscopic model \cite{pasta2}, and has a large peak near
   $w=0$.    The dotted and dashed curves are for a single component pure system using a time step $\Delta t$ of either 50 or 250 fm/c (see text).} 
\label{Fig9}
\end{center}
\end{figure}

In this subsection we calculate the dynamical response function $S(q,w)$ to study
further the difference between pure systems and systems with mixtures of ions.  The
dynamical response function describes the probability for a neutrino to transfer
momentum $q$ and energy $w$ to the system.  We have calculated $S(q,w)$ for a
microscopic nucleon model in ref. \cite{pasta3}.  The static structure factor is the
energy integral of $S(q,w)$,
\begin{equation}
   S(q)=\int_0^\infty S(q,w) dw.
\end{equation}
We calculate $S(q,w)$ as an integral over the density-density correlation function
$S(q,t)$ as follows,
\begin{equation}
   S(q,w)=\frac{1}{\pi}\int_0^{T_{max}} S(q,t) \cos(w t) dt,
\label{sqw}
\end{equation}
where $S(q,t)$ is,
\begin{equation}
   S(q,t)=\frac{1}{N_{ion} T_{ave}} \int_0^{T_{ave}} \rho(q,t+s)^*\rho(q,s) ds.
\label{sqt}
\end{equation}
Here $\rho(q,t)$ is $\hat\rho(q)$, Eq. (\ref{rhohat}), evaluated with ion coordinates
$r_i(t)$ at time $t$.  We note that $S(q)=S(q,t=0)$.  Figure \ref{Fig9} shows $S(q,w)$
for $q=0.116$ fm$^{-1}$ at a density of $1.66\times 10^{13}$ g/cm$^3$, $Y_p=0.2$,
and $T=1$ MeV, assuming either a mixture of ions from the microscopic model or a
single ion species.  All simulations show a large peak near $w=0.003$ fm$^{-1}$
(0.6 MeV) that corresponds to plasma oscillations of the ions \cite{Fe71}.  This
peak is virtually identical for calculations with a mixture of ions or a single species.
Plasma oscillations may depend on the ratio of charge density to average ion mass,
since the restoring force depends on the charge density while the oscillation frequency
also depends on one over the ion mass.  However plasma oscillations do not appear to
depend on the dispersion in the ion charge or mass distributions.

In addition, the simulation for a mixture of ions shows a large peak at $w=0$. This
peak is absent in simulations with a single ion species.  We conclude that this $w=0$
peak is the primary difference between mixtures and a single species.  The static
structure factor of a mixture is larger than $S(q)$ for a pure system by the area
under this $w=0$ peak.  Fluctuations in weak charge density, at fixed charge density,
feel no electrostatic restoring force.  Therefore these fluctuations will diffuse
slowly throughout the system and contribute to the response at low $w$.  However,
such fluctuations in weak charge concentration are only possible for mixtures.  In
a pure system the weak charge density must be proportional to the electromagnetic
charge density and there can be no weak charge fluctuations without electrostatic
restoring forces.  Therefore the response of the pure system has no large peak at $w=0$.   

The peak at $w=0$ is also seen in the microscopic calculations of Ref. \cite{pasta3}.
Here molecular dynamics simulations were preformed directly in the nucleon coordinates.
Nucleons dynamically formed clusters (nuclei) of different $N$ and $Z$ under the
influence of nuclear and Coulomb interactions.  Fluctuations in concentration of
these clusters can then diffuse as in our ion simulations.  In addition, these
microscopic calculations have two effects that are not in the ion simulations.  First,
the nuclei have a variety of excited internal modes.  Second, nucleons can diffuse
into and out of clusters leading to nuclear reactions and the changing of $N$ and $Z$.
These microscopic degrees of freedom may be responsible for significantly broadening
the plasma oscillation peak.  However, since both the ion and microscopic simulations
have the $w=0$ peak, we conclude that these microscopic degrees of freedom are not
necessary for the low energy mode.    

Finally, we have calculated $S(q,w)$ for a pure system by approximating the integrals
in Eqs. (\ref{sqw},\ref{sqt}) by a sum using a time step $\Delta t$ of 250 or 50 fm/c.
In both cases the ion trajectories have been integrated with a time step of 50 fm/c.
At high $w$ near 0.005 fm$^{-1}$  the calculation with $\Delta t=250$ fm/c shows a small
but nonzero value.  This appears to be a numerical artifact because the response for
$\Delta t=50$ fm/c is much smaller.  However, the response at $w=0$ does not show a
peak for either value of $\Delta t$.  We conclude that the large $w=0$ peak seen in
the mixture simulation, but not seen in either pure simulation, is unlikely to be a
numerical artifact.

\subsection{Finite Size Effects}
\label{subsec:finitesize}

\begin{figure}[ht]
\begin{center}
\includegraphics[width=2.75in,angle=270,clip=true] {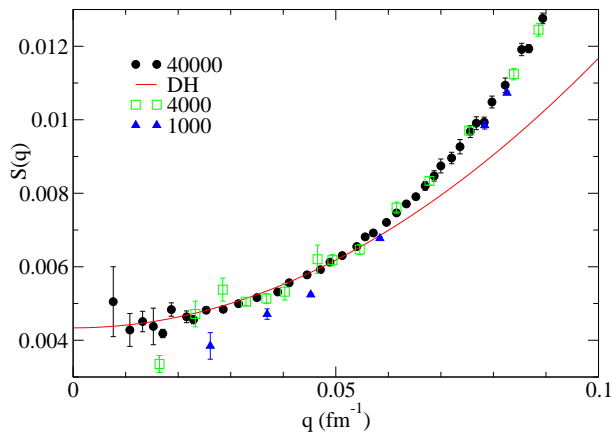}
\caption{Static structure factor $S(q)$ versus momentum transfer $q$ for conditions as in Fig. \ref{Fig5} assuming a mixture of ions from a microscopic model \cite{pasta2}.  Molecular dynamics results for 1000 ions are solid triangles, 4000 ions are open squares, and results for 40000 ions are filled circles.  The solid curve is the Debye Huckel approximation.} 
\label{Fig10}
\end{center}
\end{figure}

In this subsection we study the dependence of the MD results on the number of ions
in the simulation.  In figure \ref{Fig10} we compare $S(q)$ results for simulations
using 1000, 4000, and 40000 ions.   All simulations are for a mixture of ions at a
density of $1.66\times 10^{13}$ g/cm$^3$, as in Fig. \ref{Fig5}.  For momentum transfers
below $q=0.055$ fm$^{-1}$ there is good agreement between MD simulation results for
40000 ions and the Debye Huckel approximation.  However, despite measuring for the
relatively long time of $1.3\times 10^7$ fm/c, see run DB1 in Table \ref{tabletwo},
there are still significant statistical errors at the lowest $q$ values.  A MD simulation
with 4000 ions, see run CH1 of Table \ref{tabletwo}, agrees with the Debye Huckel $S(q)$
(and with run DB1) for $0.02 < q < 0.055$ fm$^{-1}$.  However, the smallest $q$ point
near 0.017 fm$^{-1}$ is somewhat lower.  Finally, a simulation with 1000 ions, see run
LC1 of Table \ref{tabletwo}, underpredicts the Debye Huckel $S(q)$ for $q<0.05$ fm$^{-1}$.

In order to minimize finite size effects, we calculate $S(q)$ for $q$ greater than or
equal to a minimum value $q_{min}=2\pi/L$ with $L$ the box size.  This minimum $q$ is
clearly smaller for simulations with more ions.  However, the relatively large statistical
errors found for $S(q)$ at small $q$ may be because it takes a long time for concentration
fluctuations to diffuse across a large box size $L$.  Therefore, in order to simulate
$S(q)$ accurately at small $q$ it appears to be necessary to both use a large number of
particles and to measure for a long time.  Nevertheless, we conclude from Fig. \ref{Fig10}
that molecular dynamics simulation results for $S(q)$ do converge to the Debye Huckel
approximation for small $q$ and that this convergence is clearly seen in simulations
with 4000 or 40000 ions and for $q<0.055$ fm$^{-1}$.

\begin{figure}[ht]
\begin{center}
\includegraphics[width=2.75in,angle=270,clip=true] {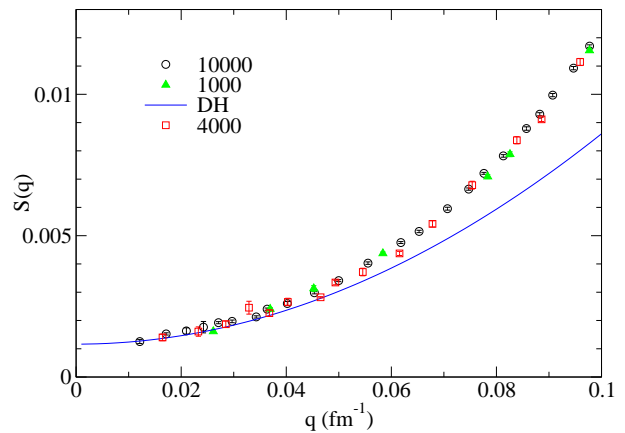}
\caption{Static structure factor $S(q)$ versus momentum transfer $q$ for conditions as in Fig. \ref{Fig5} assuming a single ion species.  Molecular dynamics results for 1000 ions (triangles), 4000 ions (squares) and 10000 ions (circles) are shown.    The solid curve is the Debye Huckel approximation. } 
\label{Fig11}
\end{center}
\end{figure}

Figure \ref{Fig11} shows $S(q)$ for a pure system composed of a single ion species
from MD simulations using 1000, 4000, and 10000 ions, see the LC2, CH2, and DB2 runs
respectively in Table \ref{tabletwo}.  Finite size effects appear smaller than in
Fig. \ref{Fig10}.  For a pure system there are no fluctuations in concentration.
Therefore difficulties with large diffusion times and statistical errors both appear
reduced at small $q$.  Again the MD results appear to converge to the Debye Huckel
approximation, although this convergence may only occur at smaller $q<0.03$ fm$^{-1}$
compared to the $q<0.055$ fm$^{-1}$ of Fig. \ref{Fig10}.

The agreement between our MD results and the Debye Huckel approximation at low $q$
provides a significant test of our simulation codes.  Furthermore three independent
MD codes CH (used for runs CH1 through CH4 of Table \ref{tabletwo}, etc.) , DB, and
LC were used.  There appears to be good agreement between the codes. 

\subsection{Neutrino Mean Free Path}
\label{subsec:mfp}
\begin{figure}[ht]
\begin{center}
\includegraphics[width=2.75in,angle=270,clip=true] {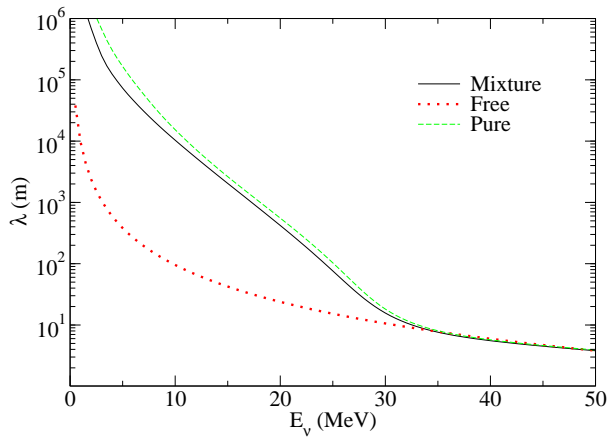}
\caption{Neutrino transport mean free path $\lambda$ versus neutrino energy $E_\nu$
   at a density of $1.66\times 10^{13}$ g/cm$^3$.  The solid curve shows results for
   a mixture of ions while the dashed curve is for a pure single species of ion.
   Finally, the dotted curve is for free ions and neglects ion-ion correlations.  } 
\label{Fig12}
\end{center}
\end{figure}

In this subsection we use our $S(q)$ results to calculate neutrino scattering mean
free paths.  Figure \ref{Fig12} shows the neutrino transport mean free path $\lambda$
(for nucleus elastic scattering) versus neutrino energy at a density of $1.66\times
10^{13}$ g/cm$^3$ based on simulations CH1, and CH2 of Table \ref{tabletwo}.  We see
that ion-ion correlations significantly increase $\lambda$ compared to that for free
ions.  

\begin{figure}[ht]
\begin{center}
\includegraphics[width=2.75in,angle=270,clip=true] {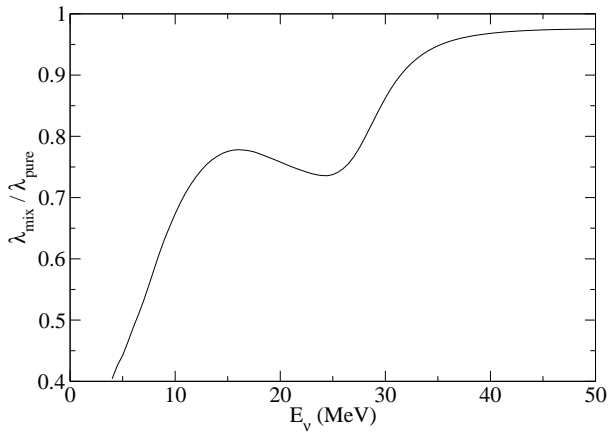}
\caption{Ratio of neutrino transport mean free paths versus neutrino energy $E_\nu$
   for a mixture of ions compared to that for a pure system.  Conditions as in Fig.
   \ref{Fig12}. } 
\label{Fig13}
\end{center}
\end{figure}

Next, Fig. \ref{Fig13} shows the ratio of mean free paths from Fig. \ref{Fig12} for
a mixture of ions compared to a single ion species.  The mean free path for a mixture
can be up to a factor of two shorter than that for a pure system.
  
\section{Summary and Conclusions}
\label{sec:conclusions}
Neutrinos in core collapse supernovae are likely first trapped by large neutrino-nucleus
elastic scattering cross sections.  These cross sections are reduced by ion-ion
correlations.  In this paper, we present classical molecular dynamics (MD) simulations
of ion systems with strong Coulomb interactions.  We find that neutrino cross sections
for mixtures of ions, with a dispersion in the ratio of neutron to proton number, are
systematically larger than those for a pure system composed of a single ion species.
We consider ion compositions from both a microscopic dynamical model and a statistical
model.  

To investigate the difference between ion mixtures and pure systems we calculate the
dynamical response function $S(q,w)$.  This describes the probability for a neutrino
to transfer momentum $q$ and energy $w$.  We find that mixtures have an extra peak in
$S(q,w)$ at $w=0$ that corresponds to diffusion of composition fluctuations.  This
peak is absent in simulations of a single ion species.

Our exact MD simulation results for the static structure factor $S(q)$ reduce to the
Debye Huckel approximation in the limit of small momentum transfer $q$.  However, this
reduction may only happen at very small $q$.    We have studied finite size effects by
comparing MD simulations with 1000 to 40000 ions.  Finite size effects appear small for
4000 or more ions.  However, it may be necessary to measure $S(q)$ for long simulation
times in order to obtain accurate  results at small $q$.

Neutrino transport mean free paths can be as much as a factor of two shorter in mixtures
compared to a pure system, for low neutrino energies and neutron rich conditions.  The
dispersion in composition may be smaller in less neutron rich systems.  Finally, an
important remaining uncertainty is simply the average ion composition predicted by
different models.  The neutrino mean free path can be significantly influenced by changes
in the average nuclear size.

\section{Acknowledgements}
\label{ack}
We thank A. Botvina for providing composition results from his statistical model and
R. Sawyer for useful discussion.  This work was supported in part by DOE grant
DE-FG02-87ER40365 and by Shared University Research grants from IBM, Inc. to Indiana
University.

\vfill\eject


\begin{thebibliography}{99} 
\bibitem{ref1} S. Woosley and H.-T. Janka, Nature Physics {\bf 1} (2005) 147.
\bibitem{ref1a} A. Burrows, Ann. Rev. Nuc. and Part. Sci. {\bf 40} (1990) 181.
\bibitem{ref2}D. Z. Freedman, D. N. Schramm, and D. L. Tubbs, Ann. Rev. Nucl. Sci. {\bf 27} (1977) 167.
\bibitem{ecap1} W. R. Hix et al., Phys. Rev. Lett. {\bf 91} (2003) 201102.
\bibitem{ecap2} S. W. Bruenn, A. Mezzacappa, Phys. Rev. D {\bf 56} (1997) 7529.
\bibitem{Ho97} C.J. Horowitz,
              Phys. Rev. D ~{\bf 55}, 4577 (1997).
\bibitem{itoh} 
N. Itoh, R. Asahara, N. Tomizawa, S. Wanajo, and S. Nozawa, Astrophys.J. {\bf 611} (2004) 1041.
\bibitem{janka} A. Marek, H.-Th. Janka, R. Buras, M. Liebendoerfer, M. Rampp, astro-ph/0504291.  
\bibitem{escreen1} L. B. Leinson, V. N. Oraevsky, and V. B. Semikoz, Phys. Lett. {\bf B209} (1988) 80.
\bibitem{escreen2} C. J. Horowitz and K. Wehrberger, Phys. Rev. L. {\bf 66} (1991) 272.
\bibitem{sawyer} R. F. Sawyer, Phys. Lett. {\bf B630} (2005) 1.
\bibitem{botvina} A. S. Botvina and I. N. Mishustin,  Phys. Lett. B {\bf 584} (2004) 233.
\bibitem{pasta} C. J. Horowitz, M. A. Perez-Garcia, and J. Piekarewicz, 
	Phys.Rev. {\bf C69} (2004) 045804.
 \bibitem{pasta2} C.J. Horowitz, M.A. Perez-Garcia, J. Carriere, D. K. Berry, and J. Piekarewicz,      	Phys.Rev. {\bf C70} (2004) 065806.
\bibitem{No64} P. Nozieres,
             Theory of Interacting Fermi Systems (Benjemin, N.Y., 1964),p.53 
\bibitem{Fe71} A. L. Fetter and J. D. Walecka,
               Quantum Theory of Many Body Systems (McGraw-Hill, New York,1971).
\bibitem{Er97} F. Ercolessi, A Molecular Dynamics Primer,
               available from http://www.sissa.it/furio/ (1997).
\bibitem{pasta3}C. J. Horowitz, M.A. PŽrez-Garc'a, D. K. Berry, J. Piekarewicz,
	 Phys.Rev. {\bf C72} (2005) 035801.



\end{thebibliography}
\end{document}